\documentclass[twocolumn,english]{revtex4-2}
\usepackage[T1]{fontenc}
\usepackage[latin9]{inputenc}
\usepackage[a4paper]{geometry}
\usepackage{amsmath,amssymb,amsfonts}
\usepackage{comment}
\usepackage{kbordermatrix}
\renewcommand{\kbldelim}{(}
\renewcommand{\kbrdelim}{)}

\begin{document}
\widetext

\title{Yangian Symmetry in Five Dimensions}
\author{Arthur Lipstein $^{a}$}
\author{Tristan Orchard $^{b}$}
\affiliation{$^{a}$Department of Mathematical Sciences, Durham University, Durham, DH1 3LE, UK}
\affiliation{$^{b}$Department of Mathematics, King's College London, The Strand, WC2R 2LS, UK}

\begin{abstract}
Quantum gravity in AdS$_7 \times$S$^4$ is dual to a 6d superconformal field theory, known as the 6d $(2,0)$ theory, which is very challenging to describe because it lacks a conventional Lagrangian description. On the other hand, certain null reductions of the 6d $(2,0)$ theory give rise to 5d Lagrangian theories with $SU(1,3)$ spacetime symmetry, $SO(5)$ R-symmetry, and 24 supercharges. This appears to be closely related to the superconformal group of a 3d superconformal Chern-Simons theory known as the ABJM theory, which is believed to be integrable in the planar limit, if one exchanges the role of conformal and R-symmetry. In this note, we construct a representation of the 5d superconformal group using 6d supertwistors and show that it admits an infinite dimensional extension known as Yangian symmetry, which opens up the possiblity that these 5d theories are exactly solvable in the planar limit. 
\end{abstract}

\maketitle

\section{Introduction}

One of the cornerstones of modern physics is the concept of symmetry, which constrains the form of particle interactions and provides powerful theoretical tools for computing observables. For example,  Lorentz and gauge symmetry are fundamental ingredients of the Standard Model while conformal symmetry is essential for describing critical phenomena and quantum gravity in Anti-de Sitter space via the AdS/CFT correspondence. Moreover supersymmetry plays a prominent role in string theory and various possible extensions of the Standard Model. The three canonical examples of AdS/CFT relate superconformal theories in three, four, and six spacetime dimensions to quantum gravity in AdS$_4\times$S$^7$, AdS$_5\times$S$^5$, and AdS$_7\times$S$^4$, respectively \cite{Maldacena:1997re}. The superconformal theories in the first two cases (known as the ABJM theory \cite{Bagger:2007jr,Bagger:2007vi,Gustavsson:2007vu,Aharony:2008ug} and $\mathcal{N}=4$ super Yang-Mills \cite{Brink:1976bc}) are well-understood, but the third case (known as the 6d $(2,0)$ theory) remains very mysterious despite decades of research. 

In the planar limit, $\mathcal{N}=4$ super Yang-Mills and the ABJM theory have a remarkable property known as integrability, which makes it possible to explicitly compute many observables to all orders in perturbation theory (see \cite{Beisert:2010jr} and references therein). A hallmark of integrability is the presence of an infinite dimensional symmetry known as Yangian symmetry. The Yangian algebra is a graded algebra whose level-0 generators correspond to the superconformal groups $PSU(2,2|4)$ and $OSp(6|4)$, in $\mathcal{N}=4$ super Yang-Mills and the ABJM theory, respectively. Since integrability is usually restricted to two-dimensional models, the above property hints at their secret equivalence to string theory, which is described by a two-dimensional sigma model (more precisely, the ABJM theory is only described by string theory in a certain limit; at strong coupling it is dual to M-theory). Non-supersymmetric integrable theories known as conformal fishnet theories can also be obtained by deforming $\mathcal{N}=4$ super Yang-Mills and the ABJM theory \cite{Gurdogan:2015csr,Caetano:2016ydc,Gromov:2017cja}. A conformal fishnet theory has also been found in 6d, although the underlying superconformal theory is unknown \cite{Chicherin:2017frs}.

Given the above considerations, it is natural to wonder if the 6d $(2,0)$ theory, which has superconformal group $OSp(8|4)$, exhibits some form of integrability as well. While this theory is not believed to have a conventional Lagrangian description, such descriptions can be obtained by dimensional reduction to five dimensions \cite{Douglas:2010iu,Lambert:2010iw,Lambert:2011gb,Kim:2012tr,Hull:2014cxa}. In this paper, we will consider a class of Lagrangians which describe a particular null reduction of the 6d $(2,0)$ theory which breaks the 6d conformal group to $SU(1,3)\times U(1)$ while preserving the R-symmetry and three quarters of the superconformal symmetry \cite{Lambert:2019jwi,Lambert:2019fne}. The spacetime symmetry contains a Lifshitz scaling and can therefore be thought of as a non-relativistic conformal symmetry. The Lagrangians are 5d $\Omega$-deformed super Yang-Mills theories with a Lagrange multiplier which localises the path integral onto anti-self-dual field configurations. The instanton solutions and correlation functions of these theories exhibit rich mathematical structure, from which many observables of the 6d theory can in principle be derived \cite{Lambert:2020zdc,Lambert:2021mnu,Lambert:2021fsl}. 

Note that the 5d superconformal group described above appears to be closely related to that of the ABJM theory if one exchanges the roles of conformal and R-symmetry and performs Wick-rotations. As a first step towards understanding the possible role of integrability in the 6d $(2,0)$ theory, we therefore ask the following group theoretic question: does the 5d superconformal group admit a Yangian extension? In this paper we answer this question in the affirmative. Since the Killing form vanishes we must find a representation with a nontrivial bilinear form in order to construct the Yangian. Using supertwistors of the 6d superconformal group, we construct bi-local operators corresponding to level-1 generators and verify that they obey a deformed analogue of the Jacobi relations known as the Serre relations, which is a necessary and sufficient condition to have a Yangian algebra. Since the representation we use is not the fundamental representation of the 5d superconformal group, standard theorems guaranteeing that the Serre relations are satisfied do not apply, making our construction very nontrivial.  

This result is significant for several reasons. First of all, it represents the first example of a Yangian extension of a superconformal group in five dimensions. It also represents the first example based on a non-relativistic conformal group above two dimensions (for recent work on non-relativistic integrable theories in two dimensions see \cite{Bastianello:2016rra,Harmark:2018cdl,Harmark:2020vll}). Furthermore, if this turns out to be a symmetry of the 5d $\Omega$-deformed super Yang-Mills theories described above, this would strongly suggest that they are exactly solvable in the planar limit and would provide a powerful new toolbox for analyzing 5d gauge theories, which are typically non-renormalisable and therefore far less understood than lower dimensional gauge theories. Finally, it would have important implications for quantum gravity via the AdS/CFT correspondence, as we discuss in the conclusion.

\section{Null Reduction of 6d $(2,0)$}

While the interacting 6d $(2,0)$ theory does not have a conventional Lagrangian description, much can be learned from dimensional reduction. Let us therefore consider the $(2,0)$ theory on a manifold with metric
\begin{align}
ds^2 = -2dx^+(dx^- - \frac12 \Omega_{ij}x^idx^j) + dx^idx^i,
\label{metric}
\end{align}
where $i \in \left\{ 1,2,3,4\right\} $, $-\pi R \leq x^+ \leq \pi R$, and $\Omega$ is an anti-self-dual 2-form satisfying $\Omega_{ik} \Omega_{jk}=R^{-2}\delta_{ij}$. This metric can be obtained from a standard 6d Minkowski metric $ds^2 =d\hat{x}^{\mu} d\hat{x}_{\mu}$ via a change of variables and Weyl transformation \cite{Lambert:2020zdc} and describes the boundary of AdS$_7$ with radius $R$ when written as a $U(1)$ fibration over a non-compact complex projective space \cite{Pope:1999xg,Lambert:2019jwi}. Note that the Weyl anomaly of the 6d $(2,0)$ theory \cite{Henningson:1998gx} vanishes for the metric in \eqref{metric} \cite{Lambert:2019jwi}.  

The field content of the abelian 6d $(2,0)$ theory consists of a self-dual 2-form, 5 scalar fields, and 8 Majorana-Weyl fermions \cite{Howe:1983fr,Perry:1996mk,Pasti:1997gx}. Reducing it along $x^+$ and non-abelianising gives the following Lagrangian theory \cite{Lambert:2019jwi}:
\begin{align}
\begin{split}
    \mathcal{L}= \frac{k}{4\pi^2 R} \text{Tr} \Big\{ \frac{1}{2} F_{i -} F^{i}{}_{-} + \frac{1}{2}  G_{ij} \mathcal{F}^{ij} - \frac{1}{2} \nabla_{i} X^I \nabla^i X^I  \\ + \frac{i}{2} \bar{\psi}_A\Gamma^{-}D_{-} \psi^A +\frac{i}{2}  \bar{\psi}_A \Gamma^{i} \nabla_i \psi^A  + \frac{1}{2} \bar{\psi}_A \Gamma_{+} \widetilde{\Gamma}^{I} \big [ X^I, \psi^A \big ]  \Big\},
    \end{split}
\label{lag}
\end{align}
where $\nabla_i = D_i - \frac{1}{2} \Omega_{i j} x^{j} D_-$, with $D_-$ and $D_i$ being standard covariant derivatives for the gauge fields $A_-$ and $A_i$, $G$ is a self-dual Lagrange multiplier, $\mathcal{F}$ is an anti-self-dual field strength constructed from a linear combination of the the field strengths $F_{ij}$ and $F_{-i}$, and $k$ is an orbifold parameter \cite{Lambert:2020zdc,Lambert:2021fsl}. The $X^I$, where $I\in\left\{ 1,...,5\right\}$, are scalars transforming in the $\mathbf{5} $ of the R-symmetry group $SO(5)$, while the spinors are 6d symplectic-Majorana-Weyl transforming in the $\mathbf{4}$ of  the $USp(4)\backsimeq SO(5)$ R-symmetry, i.e. $A \in\left\{ 1,...,4\right\} $. Our spinor conventions are summarised in Appendix B. All fields are valued in the adjoint of $SU(N)$. Setting $\Omega=0$ provides a field theoretic realisation of the DLCQ description of the 6d $(2,0)$ theory \cite{Aharony:1997an}. The t'Hooft coupling is given by $\lambda = g_{YM}^2 N  \propto N R / k$. The planar limit can then be defined by taking $N$ and $k$ to infinity while holding $\lambda$ fixed. Moreover in the regime $N \ll k^3$ we expect the gravitional dual to be IIA string theory on $\tilde{\mathbb{CP}}^3\times S^4$ (where $\tilde{\mathbb{CP}}^3$ is a non-compact complex projective space) \cite{Lambert:2019jwi}.

\section{Superconformal Symmetry} \label{symmetry}
The non-relativistic superconformal symmetries of the 5d theory in \eqref{lag} are generated by the maximal subalgebra of the 6d superconformal group $OSp(8|4)$ that commute with $P_+$, the generator of $x^+$ translations which are an isometry of the metric in \eqref{metric}. Denoting the superconformal generators of the 6d $(2,0)$ theory in Minkowski background with hatted indices, we find that $P_+$ is given by \cite{Lambert:2020zdc} 
\begin{equation}
P_+	= \hat{P}_{+} + \frac{1}{4} \Omega_{ij} \hat{M}_{ij} + \frac{1}{8R^2}\hat{K}_{-}.
\label{pplus}
\end{equation}
The bosonic generators of the 5d subalgebra are then given by
\begin{gather}
P_-	= \hat{P}_{-}, \quad P_i= \hat{P}_i + \frac{1}{2} \Omega_{ij} \hat{M}_{j-}, \quad K_+ =	\hat{K}_+, \nonumber \\
T =	\hat{D}-\hat{M}_{+-}, \quad M_{i+} =	\hat{M}_{i+}- \frac{1}{4} \Omega_{ij} \hat{K}_j, \nonumber \\
C^\kappa	 =\frac{1}{4}\eta^\kappa_{ij} \hat{M}_{ij}, \quad B =\frac{1}{2} R \, \hat{P}_{+} - \frac{1}{8} R \, \Omega_{ij} \hat{M}_{ij} + \frac{1}{16 R}\hat{K}_{-}, 
\label{spacetimesymm}
\end{gather}
where $\eta^\kappa_{ij}$ are 4d self-dual t'Hooft matrices with $\kappa\in\left\{ 1,2,3\right\} $ and the R-symmetry generators are
\begin{align}
R_{IJ} &= \hat{R}_{IJ},
\label{rsymm}
\end{align}
where $R_{IJ}$ is symmetric and traceless with respect to the invariant tensor of $USp(4)$. Hence, the 6d conformal symmetry is broken from $SO(2,6)$ to $SU(1,3) \times U(1)$, where the $U(1)$ is generated by $P_+$, and the $USp(4)$ R-symmetry is unbroken. Note that $T$ generates is a Lifshitz scaling $\left(x^{-},x^{i}\right)\rightarrow\left(\Lambda^{2}x^{-},\Lambda x^{i}\right)$, so this can be thought of as a non-relativistic conformal group. 

We also find the following fermionic generators \cite{Lambert:2019jwi,Lambert:2021nol}:
\begin{align}
Q_{-} &= \Gamma_{-} \hat{Q}  \nonumber \\  
S_{+} &= \Gamma_{+} \hat{S}  \nonumber \\
\Theta_{-} &= \frac{1}{4} \Big( R \, \Omega_{ij} \Gamma_{ij} \hat{Q} + \frac{1}{R} \Gamma_{-} \hat{S} \Big). \label{5dfermion2}
\end{align}
The hatted spinors are 6d symplectic-Majorana-Weyl spinors in the $\mathbf{4}$ of the $USp(4)$ R-symmetry (for simplicity, we have suppressed their spinor and R-symmetry indices). This gives 16 real components for $\hat{Q}$ and $\hat{S}$. The above combinations of $\Gamma$-matrices act as projectors, reducing this by half so $Q_{-}, S_{+}$ and $\Theta_{-}$ each have 8 real components, with the subscripts indicating the chirality under $\Gamma_{+-}$. Hence there are are total of 24 superconformal charges. 

The 5d superconformal algebra can be deduced from the 6d one and one finds that the bosonic subalgebra generates $U(1) \times SU(1,3) \times SO(5) \, \cong \, U(1) \times SO(6) \times USp(4)$ (see Appendices C and D for more details). We will later require a non-degenerate bilinear form to raise and lower adjoint indices. The natural choice is the Killing form, which is defined in terms of the adjoint representation:
\begin{align}
g_{\text{Ad}}^{m n} = \text{Str} \big( \text{Ad} (J^m) \cdot \text{Ad} (J^n) \big) = \sum_{p, \, q} (-1)^{|q|} f^{m p}_{\ \ q} f^{n q}_{\ \ p}.  
\end{align}
It is a standard result of Lie superalgerbas that the Killing form vanishes for $OSp(2n+2|2n)$, and we have explicitly checked this for the 5d superalgebra. We will therefore need to use a different representation, which we describe in the next section.

\section{Twistorial Representation} \label{twistorrep}
Since the Killing form of the 5d superalgebra generated by \eqref{pplus}-\eqref{5dfermion2} vanishes, we must use a different representation in order to define a non-degenerate bilinear form on the superalgebra. We will construct it from the twistorial representation of the 6d superconformal symmetry group $OSp(8|4)$ \cite{Huang:2010rn}:
\begin{equation}
\mathcal{Z}^{\mathcal{M} \hat{a}}=\left(\lambda^{\hat{\alpha}\hat{a}},\mu_{\hat{\beta}}^{\hat{a}},\eta^{\hat{A} \hat{a}},\tilde{\eta}^{\hat{a}}_{\hat{B}}\right).\label{eq:twistor}
\end{equation}
The indices $\hat{\alpha},\hat{\beta}\in\left\{ 1,2,3,4\right\} $ correspond to chiral spinor indices of the 6d Lorentz group $SU(4)$, while $\hat{a}\in\left\{ 1,2\right\} $ label the fundamental representation of $SU(2)$ which arises from the 6d little group $SO(4)=SU(2)\times SU(2)$. These indices are raised and lowered using the two-index antisymmetric tensor $\varepsilon_{\hat{a}\hat{b}}$ and its inverse. $\hat{A},\hat{B}\in\left\{ 1,2\right\} $ label the fundamental representation of a subgroup of the R-symmetry $U(2)\subset USp(4)$. This subgroup arises from using harmonic superspace variables parameterising the coset $\frac{USp(4)}{U(1)\times U(1)}$ \cite{Ferrara:1999bv}. Our index conventions are summarised in Appendix A. Note that $\left(\lambda,\mu\right)$ are bosonic and $\left(\eta,\tilde{\eta}\right)$ are fermionic. Supertwistors have also been used to study Yangian symmetry of $\mathcal{N}=4$ super Yang-Mills \cite{Drummond:2009fd} and the ABJM theory \cite{Bargheer:2010hn}. 

In terms of the variables in \eqref{eq:twistor}, the 6d superconformal generators are given by
\begin{gather}
\hat{P}^{\hat{\alpha}\hat{\beta}}=\lambda^{\hat{\alpha}\hat{a}}\lambda_{\hat{a}}^{\hat{\beta}},\quad \hat{K}_{\hat{\alpha}\hat{\beta}}=\frac{\partial^{2}}{\partial\lambda^{\hat{\alpha}\hat{a}}\partial\lambda_{\hat{a}}^{\hat{\beta}}}, \nonumber \\
\hat{D}=\frac{1}{2}\lambda^{\hat{\alpha}\hat{a}}\partial_{\lambda^{\hat{\alpha}\hat{a}}}+2, \quad 
\hat{M}_{\ \hat{\beta}}^{\hat{\alpha}}=\lambda^{\hat{\alpha}\hat{a}}\partial_{\lambda^{\hat{\beta}\hat{a}}}-\frac{1}{4}\delta_{\hat{\beta}}^{\hat{\alpha}}\lambda^{\hat{\gamma} \hat{a}}\partial_{\lambda^{\hat{\gamma} \hat{a}}} \nonumber \\ 
\hat{R}^{\hat{A}\hat{B}}=\eta^{\hat{A}\hat{a}}\eta_{\hat{a}}^{\hat{B}},\,\,\,\hat{R}_{\hat{A}\hat{B}}=\frac{\partial^{2}}{\partial\eta^{\hat{A}\hat{a}}\partial\eta_{\hat{a}}^{\hat{B}}},\,\,\,\hat{R}_{\hat{B}}^{\hat{A}}=\eta^{\hat{A}\hat{a}}\partial_{\eta^{\hat{B}\hat{a}}}-\delta_{\hat{B}}^{\hat{A}} \nonumber \\
\hat{Q}^{\hat{\alpha}\hat{A}}=\lambda^{\hat{\alpha}\hat{a}}\eta_{\hat{a}}^{\hat{A}}, \quad \hat{Q}_{\hat{A}}^{\hat{\alpha}}=\lambda^{\hat{\alpha}\hat{a}}\partial_{\eta^{\hat{A}\hat{a}}}, \nonumber \\
\hat{S}_{\hat{\alpha} \hat{A}}=\frac{\partial^{2}}{\partial\lambda^{\hat{\alpha}\hat{a}}\partial\eta_{\hat{a}}^{\hat{A}}}, \quad \hat{S}_{\hat{\alpha}}^{\hat{A}}=\eta^{\hat{A} \hat{a}}\partial_{\lambda^{\hat{\alpha} \hat{a}}}.
\label{6dgeneratorstwistors}
\end{gather}
Note that this representation is not linear. A linear representation can be obtained by noting that the supertwistors are self-conjugate 
\begin{equation}
\lambda^{\hat{\alpha}\hat{a}}=\partial_{\mu_{\hat{\alpha}\hat{a}}},\,\,\,\mu_{\hat{\alpha}\hat{a}}=-\partial_{\lambda^{\hat{\alpha}\hat{a}}},\,\,\,\eta^{\hat{A}\hat{a}}=-\partial_{\tilde{\eta}_{\hat{A}\hat{a}}},\,\,\,\tilde{\eta}_{\hat{A}\hat{a}}=-\partial_{\eta^{\hat{A}\hat{a}}},\label{relations}
\end{equation}
and using these relations to extend the action of the superconformal generators in \eqref{6dgeneratorstwistors} to $\mu$ and $\tilde{\eta}$. For example,
writing the momentum generator as a linear operator acting on both $\lambda$ and $\mu$
gives 
\begin{equation}
\hat{P}^{\hat{\alpha}\hat{\beta}}=\varepsilon_{\hat{a} \hat{b}}\left(\lambda^{\hat{\alpha}\hat{a}}\partial_{\mu_{\hat{\beta}\hat{b}}}-\lambda^{\hat{\beta}\hat{a}}\partial_{\mu_{\hat{\alpha}\hat{b}}}\right).
\label{linearexample1}
\end{equation}
Similarly, the special conformal symmetry generator $\hat{S}_{\hat{\alpha}\hat{A}}$ can be written as a linear operator as follows: 
\begin{equation}
\hat{S}_{\hat{\alpha}\hat{A}}=\varepsilon^{\hat{a}\hat{b}}\left(-\mu_{\hat{\alpha}\hat{a}}\partial_{\eta^{\hat{A}\hat{b}}}+\tilde{\eta}_{\hat{A}\hat{a}}\partial_{\lambda^{\hat{\alpha}\hat{b}}}\right).
\label{linearexample2}
\end{equation}

In this way the 6d superconformal generators can be written as $(16+8)\times(16+8)$ supermatrices which can be readily implemented on a computer and used to check the Yangian algebra, as we explain in the next section and Appendix E. We denote the matrix representation by $\mathcal{R}$ and spell it out below:
\begin{widetext}
\begin{align}
\label{eq:MatrixRep}
\begin{aligned}
&\mathcal{R} \begin{bmatrix}
\begin{pmatrix}
\hat{M}^{\hat{\alpha}}_{\ \hat{\beta}}	& \hat{P}^{\hat{\alpha} \hat{\beta}}		& \vline & \hat{Q}^{\hat{\alpha}}_{\ \hat{A}}& \hat{Q}^{\hat{\alpha} \hat{A}}\\
\hat{K}_{\hat{\alpha} \hat{\beta}}	 	& \hat{M}_{\hat{\alpha}}^{\ \hat{\beta}}	& \vline & \hat{S}_{\hat{\alpha} \hat{A}}	 & \hat{S}_{\hat{\alpha}}^{\ \hat{A}}\\
\vspace{- 10pt}				 	&								& \vline &									 & 							\\
\hline						 	&								& \vline &									 & 							\\
\vspace{- 21pt} \\
\hat{S}^{\hat{A}}_{\ \hat{\alpha}}&\hat{Q}^{\hat{A}\hat{\alpha}}& \vline & \hat{R}^{\hat{A}}_{\ \hat{B}}	 	 & \hat{R}^{\hat{A} \hat{B}}	\\
\hat{S}_{\hat{A} \hat{\alpha}}&\hat{Q}_{\hat{A}}^{\ \hat{\alpha}}& \vline & \hat{R}_{\hat{A} \hat{B}}			 &	\hat{R}_{\hat{B}}^{\ \hat{A}}	\\
\end{pmatrix}
\end{bmatrix} =
\begin{pmatrix}
(E^{\hat{\alpha}}_{\ \hat{\beta}} - \frac{1}{4} \delta^{\hat{\alpha}}_{\hat{\beta}} \, \mathbb{I}_4 )\! \otimes \! \mathbb{I}_2	& (E^{\hat{\alpha} \hat{\beta}}-E^{\hat{\beta} \hat{\alpha}})\! \otimes \! \varepsilon_2 & \vline & E^{\hat{\alpha}}_{\ \hat{A}} \! \otimes \! \mathbb{I}_2 & - E^{\hat{\alpha} \hat{A}}\! \otimes \! \varepsilon_2 \\
(E_{\hat{\beta} \hat{\alpha}}-E_{\hat{\alpha} \hat{\beta}})\! \otimes \! \varepsilon_2^{-1} 	& (\frac{1}{4} \delta^{\hat{\beta}}_{\hat{\alpha}}\mathbb{I}_4  -E_{\hat{\alpha}}^{\ \hat{\beta}} )\! \otimes \! \mathbb{I}_2	& \vline & -E_{\hat{\alpha} \hat{A}} \! \otimes \! \varepsilon_2^{-1} & E_{\hat{\alpha}}^{\ \hat{A}} \! \otimes \! \mathbb{I}_2 \\
\vspace{- 10pt}				 	&								& \vline &									 & 							\\
\hline						 	&								& \vline &									 & 							\\
\vspace{- 21pt} \\
E^{\hat{A}}_{\ \, \hat{\alpha}} \! \otimes \! \mathbb{I}_2&-E^{\hat{A} \hat{\alpha}} \! \otimes \! \varepsilon_2 & \vline & E^{\hat{A}}_{\ \hat{B}} \! \otimes \! \mathbb{I}_2	 & \hspace*{-10pt} -(E^{\hat{A} \hat{B}} + E^{\hat{B} \hat{A}})\! \otimes \! \varepsilon_2  	\\
E_{\hat{A} \hat{\alpha}} \! \otimes \! \varepsilon_2^{-1} &- E_{\hat{A}}^{\ \, \hat{\alpha}} \! \otimes \! \mathbb{I}_2 & \vline & (E_{\hat{A} \hat{B}} + E_{\hat{B} \hat{A}})\! \otimes \! \varepsilon_2^{-1}  &	-E_{\hat{B}}^{\ \hat{A}} \! \otimes \! \mathbb{I}_2	\\
\end{pmatrix}, \\
&\hfill \\
\end{aligned}
\end{align}
\end{widetext}
\vspace{-.1in}
where $\mathbb{I}_m$ is an $m \times m$ unit matrix, $\varepsilon_2$ refers to $\varepsilon_{\hat{a}\hat{b}}$, and $E^{\rho}_{\xi}$ is a rectangular matrix with $1$ in the $\rho^{\text{th}}$ row and $\xi^{\text{th}}$ column and zeros everywhere else. The matrix elements in \eqref{eq:MatrixRep} can be read off from linear representations like \eqref{linearexample1} and \eqref{linearexample2} by noting that the rows and columns are labelled by the components of $(\lambda,\mu,\eta,\tilde{\eta})$. For notational consistency, the labels of some generators on the left-hand-side appear in a slightly different order than those in \eqref{6dgeneratorstwistors}. Dilatations are given by 
\begin{align}
\mathcal{R} \begin{bmatrix}
\begin{pmatrix}
\hat{D}	&0		&\vline	&0	&0	\\
0	 	&\hat{D}&\vline	&0	&0	\\
\vspace{-10pt}& &\vline &	&	\\
\hline	&		&\vline	&	&	\\
\vspace{-21pt} \\
0		&0		&\vline	&0	&0	\\
0		&0		&\vline	&0	&0	\\
\end{pmatrix}
\end{bmatrix} =
\begin{pmatrix}
\frac{1}{2}\mathbb{I}_8	&0		&\vline	&0	&0	\\
0	 	& \hspace*{-5pt} -\frac{1}{2}\mathbb{I}_8 &\vline	&0	&0	\\
\vspace{-10pt}& &\vline &	&	\\
\hline	&		&\vline	&	&	\\
\vspace{-21pt} \\
0		&0		&\vline	&0	&0	\\
0		&0		&\vline	&0	&0	\\
\end{pmatrix}.
\end{align}

The matrix representation for 6d generators labelled by Lorentz indices is given by
\begin{gather}
\hat{P}_{\mu} = \ \frac{1}{2} \mathcal{R}(\hat{P}^{\hat{\alpha} \hat{\beta}}) \, \widetilde{\Sigma}_{\mu \hat{\beta} \hat{\alpha}}, \quad \hat{K}_{\mu} = \ \frac{1}{2}  \mathcal{R}(\hat{K}_{\hat{\alpha} \hat{\beta}}) \, \Sigma_{\mu}^{\hat{\beta} \hat{\alpha}}, \nonumber \\
\hat{M}_{\mu \nu} = \ - \frac{1}{2}  \mathcal{R}(\hat{M}^{\hat{\alpha}}_{\ \hat{\beta}}) \, \widetilde{\Sigma}_{\mu \nu \  \hat{\alpha}}^{ \ \ \hat{\beta}}, 
\end{gather}
where $\Sigma$ and $\tilde{\Sigma}$ are Clebsch-Gordan coefficients whose precise definition in terms of the 6d Clifford algebra can be found in Appendix B. One can then obtain a matrix representation for the 5d generators in \eqref{spacetimesymm} by taking appropriate linear combinations. Moreover, the matrix representation of the 5d R-symmetry generators is simply given by
\begin{equation}
\left\{ R^{\hat{A}\hat{B}},R_{\hat{A}\hat{B}},R_{\hat{B}}^{\hat{A}}\right\} =\left\{ \mathcal{R}\left(\hat{R}^{\hat{A}\hat{B}}\right),\mathcal{R}\left(\hat{R}_{\hat{A}\hat{B}}\right),\mathcal{R}\left(\hat{R}_{\hat{B}}^{\hat{A}}\right)\right\}. 
\end{equation} 
Finally, the matrix represenation of the 5d fermionic generators is 
\begin{align}
Q_{- \hat{\alpha}}^{\hat{A}}&=\tilde{\Sigma}_{-\hat{\alpha}\hat{\beta}}\mathcal{R}(\hat{Q}^{\hat{\beta}\hat{A}}),\qquad \, Q_{- \hat{\alpha}\hat{A}}=\tilde{\Sigma}_{-\hat{\alpha}\hat{\beta}}\mathcal{R}(\hat{Q}_{\hat{A}}^{\hat{\beta}})  \nonumber \\
S_{+}^{\hat{\alpha} \hat{A}}&=\Sigma_{+}^{\hat{\alpha}\hat{\beta}}\mathcal{R}(\hat{S}_{\hat{\beta}}^{\hat{A}}),\qquad \qquad {S_{+}}_{\hat{A}}^{\! \! \! \hat{\alpha}}=\Sigma_{+}^{\hat{\alpha}\hat{\beta}}\mathcal{R}(\hat{S}_{\hat{\beta} \hat{A}}) \nonumber \\
\Theta_{-}^{\hat{\alpha}\hat{A}}&=\frac{1}{4}R\Omega_{ij}\left(\Sigma_{i}\tilde{\Sigma}_{j}\right)_{\hat{\beta}}^{\hat{\alpha}}\mathcal{R}(\hat{Q}^{\hat{\beta}\hat{A}})+\frac{1}{4R}\Sigma_{-}^{\hat{\alpha}\hat{\beta}}\mathcal{R}(\hat{S}_{\hat{\beta}}^{\hat{A}}) \nonumber \\
{\Theta_{-}}_{\hat{A}}^{\! \! \! \! \hat{\alpha}}&=\frac{1}{4}R\Omega_{ij}\left(\Sigma_{i}\tilde{\Sigma}_{j}\right)_{\hat{\beta}}^{\hat{\alpha}}\mathcal{R}(\hat{Q}_{\hat{A}}^{\hat{\beta}})+\frac{1}{4R}\Sigma_{-}^{\hat{\alpha}\hat{\beta}}\mathcal{R}(\hat{S}_{\hat{\beta}\hat{A}}).
\end{align}

Using the explicit representation constructed above, we can now define a metric on the 5d superalgebra:
\begin{align}
	g_{\mathcal{R}} (X,Y) = \text{Str} \left( \mathcal{R}(X)\cdot \mathcal{R}(Y) \right).
\end{align}
The non zero components of this metric are
\begin{align}
\label{metricg}
\hspace*{-10pt} g(P_+,P_+) &= -\frac{4}{R^2}, \qquad g(R^{\hat{A} \hat{B}}, R_{\hat{C} \hat{D}})  && \hspace*{-10pt} = -4 \big( \delta^{\hat{A}}_{\hat{C}} \delta^{\hat{B}}_{\hat{D}} + \delta^{\hat{B}}_{\hat{C}} \delta^{\hat{A}}_{\hat{D}} \big), \nonumber \\
\hspace*{-10pt} g(P_-, K_+) &= -8, \qquad \quad g(R^{\hat{A}}_{\ \hat{B}},R^{\hat{C}}_{\ \hat{D}}) && \hspace*{-10pt} = -4 \delta^{\hat{A}}_{\hat{D}} \delta^{\hat{C}}_{\hat{B}}, \nonumber \\
\hspace*{-10pt} g(P_i, M_{j+}) &= 4\Omega_{ij}, \qquad \, \, g(Q_{-}^{\hat{\alpha} \hat{A}}, S_{+ \hat{\beta} \hat{B}}) && \hspace*{-10pt} = -8 \delta^{\hat{\alpha}}_{\hat{\beta}} \delta^{\hat{A}}_{\hat{B}}, \nonumber \\
\hspace*{-10pt} g(C^\kappa, C^\rho) &= -2 \delta^{\kappa \rho} \qquad g(Q_{- \hat{A}}^{\hat{\alpha}}, S_{+ \hat{\beta}}^{\ \ \ \hat{B}}) && \hspace*{-10pt} = 8 \delta^{\hat{\alpha}}_{\hat{\beta}} \delta^{\hat{B}}_{\hat{A}}, \nonumber \\
\hspace*{-10pt} g(B,B) &= -1, \qquad \quad g(\Theta_{ - \hat{A}}^{\hat{\alpha}}, \Theta_{-}^{\hat{\beta} \hat{B}}) && \hspace*{-10pt} =  \frac{1}{\sqrt{2} R} \hat{C}^{\hat{\alpha} \hat{\beta}} \delta^{\hat{B}}_{\hat{A}}, \nonumber \\
\hspace*{-10pt} g(T,T) &= 8,
\end{align} 
where $\hat{C}^{\hat{\alpha} \hat{\beta}}$ is a charge conjugation matrix defined in Appendix B. Using the results of this section, we can now define a Yangian extension of the 5d superalgebra.

\section{Yangian} \label{yangianextension}
The Yangian $Y(\mathfrak{g})$ of a lie (super-)algebra $\mathfrak{g}$ consists of infinitely many levels, with $\text{dim}(\mathfrak{g})$ generators at each level. The infinite tower of generators can be derived from the level-0 generators $J_{(0)}^{a}$ and level-1 generators $J_{(1)}^{a}$, which obey the following supercommutation relations:
\begin{align}
[ J_{(0)}^{a}, \, J_{(0)}^{b} \} = f^{a b}_{\ \ c} J_{(0)}^{c}, \quad [ J_{(1)}^{a}, \, J_{(0)}^{b} \} = f^{a b}_{\ \ c} J_{(1)}^{c},
\label{algebra}
\end{align}
along with the Serre relations
\begin{align}
\begin{split}
&[J_{(1)}^{a},[J_{(1)}^{b}, \, J_{(0)}^{c} \} \} + \text{graded cyclic perms} \\
&  = \frac{1}{6} (-)^{|i||l| + |k| |n|} f_{ai}^{\ \ l} f_{b j}^{\ \ m} f_{c k}^{\ \ n} f^{i j k} \{ J_{(0)}^{l}, \,  J_{(0)}^{m}, \, J_{(0)}^{n} ] ,
\label{serre}
\end{split}
\end{align} 
where $|a|=0$ for bosonic generators and  $1$ for fermionic generators. For the 5d superconformal algebra we are considering, $a \in \{1,  \dots, 50\}$, with $J \in \{P_+, P_-, P_i, B, C^\kappa, T, M_{i+}, K_+, R_{\hat{A}\hat{B}},R^{\hat{A}\hat{B}}, R^{\hat{A}}_{\ \hat{B}}\}$ bosonic and $ J \in \{Q_{-}^{\hat{\alpha} \hat{A}}, Q_{- \hat{A}}^{\hat{\alpha}},  S_{+ \hat{\alpha} \hat{A}},S_{+ \hat{\alpha}}^{\ \ \ \hat{A}} , \Theta_{ - \hat{A}}^{\hat{\alpha}}, \Theta_{-}^{\hat{\alpha} \hat{A}} \}$ fermionic. Furthermore, $\{ \cdot, \cdot, \cdot ]$ denotes the graded symmetriser of three generators.

For a system of $N_s$ sites the level-0 and level-1 generators can be defined as follows:
\begin{align}
\label{level1a}
J_{(0)}^a & = \sum_{u}^{N_{s}} J_{(0) u}^a, \\ 
J_{(1)}^a &= f_{ b c}^{\ \ a} \sum_{u < v}^{N_s}  J_{(0) u}^c J_{(0) v}^b,
\label{level2}
\end{align}
where $J_{(0) u}^a$ is understood to act locally on the $u$-th site. The number of sites is defined abstractly and depends on the observable in question, for example the number of legs of a scattering amplitude or the number of operators in a correlation function. Note that the level-1 generators are bi-local. In principle, they can also contain local terms but we will not need to consider this. We provide several explicit examples of level-1 generators in Appendix F. Level-$k$ generators are then obtained by commuting $k$ level-1 generators and are $(k+1)$-local. To establish the existence of a Yangian extension of the 5d superalgebra constructed in the previous section, we must therefore verify that the relations in \eqref{algebra} and \eqref{serre} are satisfied for the definitions in \eqref{level1a} and \eqref{level2}. Remarkably, using the $24\times 24$ matrix representation constructed in the previous section, we have verified that this is indeed the case using computer algebra. We attach our Mathematica code, {\em 5dYangian.nb}, and review its structure in Appendix E.

A sufficient (but not necessary) condition for the Serre relations to hold is that the adjoint representation of the superconformal group appears once in the tensor product of the representation of the single-site level-0 generators with its conjugate \cite{Dolan:2004ps}. In our case, this condition is not satisfied since the single-site representation is constructed from the fundamental representation of $OSp(8|4)$. The Serre relations are therefore not guaranteed to hold and we must check them explicitly. That they indeed hold implies that our construction is very nontrivial.

If the single-site level-1 generators do not include local terms (as they do in our construction), the Serre relations for more than one site follow from the single-site Serre relations \cite{Dolan:2004ps}. It is therefore sufficient to check the single-site Serre relations in our case, which we have done for over one thousand randomly chosen examples. We have also verified the multi-site Serre relations for over a thousand randomly chosen examples, which provides a nontrivial check of our level-1 expressions and the underlying Mathematica code. 

\section{Discussion}

Exactly solvable quantum field theories above two dimensions serve as important toy models analogous to the harmonic oscillator and hydrogen atom in quantum mechanics. Apart from self-dual Yang-Mills and its dimenisonal reductions \cite{Ward:1977ta,Atiyah:1977pw,Corrigan:1977ma} and conformal fishnet theories, the only other examples we know of are $\mathcal{N}=4$ super Yang-Mills and the ABJM theory in the planar limit. These theories also arise in two of the three canonical examples of the AdS/CFT correspondence, suggesting that there is one more integrable quantum field theory waiting to be discovered above two dimensions.

In this note we demonstrate that the non-relativistic superconformal symmetry group of certain 5d $\Omega$-deformed gauge theories which arise from null reductions of the 6d $(2,0)$ theory can be extended to an infinite dimensional Yangian, providing the first example of such an extension in five dimensions. The key technical steps were to construct a representation of the 5d superconformal group with a non-zero bilinear form using 6d supertwistors and dimensional reduction, and to explicitly check the Serre relations using an efficient Mathematica code. This result opens up the exciting possiblity that these 5d gauge theories may be exactly solvable in the planar limit. It would also be interesting to see if our approach can make contact with other integrable theories \cite{Chicherin:2017frs} by lowering the amount of supersymmetry or including nontrivial local terms in the level-1 generators.

One way to demonstrate that the 5d gauge theories considered in this paper are indeed integrable would be to explore Yangian symmetry of the action using the strategy recently developed in \cite{Beisert:2018zxs}. A more conventional approach would be to identify observables in the 5d gauge theory that enjoy this symmetry. Since the representation constructed in section \ref{twistorrep} naturally describes scattering amplitudes in five dimensions \cite{Cheung:2009dc,Dennen:2009vk,Boels:2009bv,Huang:2010rn,Plefka:2014fta,Cachazo:2018hqa,Geyer:2018xgb,Chiodaroli:2022ssi}, these would be the most natural observables to consider. One way to deduce such amplitudes would be to construct solutions to the superconfonformal Ward identities which exhibit the required factorisation properties. It may also be possible to compute them by adapting the methods developed for self-dual Yang-Mills in \cite{Chalmers:1996rq}. Computing scattering amplitudes should also shed light on whether the 5d theories considered in this paper are renormalizable. In contrast to ordinary 5d super Yang-Mills theories, the 5d theories we consider have non-relativistic superconformal symmetry at the classical level and we expect this symmetry to persist in the quantum theory.

The next step would be to investigate if the amplitudes exhibit dual superconformal symmetry in the planar limit \cite{Drummond:2006rz,Drummond:2008vq,Brandhuber:2008pf,Huang:2010qy,Gang:2010gy}. In $\mathcal{N}=4$ super Yang-Mills and the ABJM theory this symmetry encodes level-1 Yangian generators \cite{Drummond:2009fd,Huang:2010qy}. Dual conformal structure was also recently found in the amplitudes of self-dual Yang-Mills \cite{Henn:2019mvc}. Moreover, in $\mathcal{N}=4$ super Yang-Mills dual superconformal symmetry is tied to amplitude/Wilson loop duality \cite{Alday:2007hr,Brandhuber:2007yx,Drummond:2007aua,Mason:2010yk,Caron-Huot:2010ryg} and self-duality of IIB string theory on AdS$_5 \times$S$^5$ under a certain combination of bosonic and fermionic T-duality transformations \cite{Berkovits:2008ic,Beisert:2008iq}. On the other hand, the origin of dual superconformal symmetry in the ABJM theory remains mysterious \cite{Henn:2010ps,Adam:2010hh,Bakhmatov:2010fp,Chen:2011vv,Colgain:2016gdj}. Since the superconformal symmetry of the 5d $\Omega$-deformed gauge thories considered in this paper appears to be closely related to that of the ABJM theory, this raises the tantalising possibility that these two theories are related by some analogue of T-duality \cite{Jeon:2012fn}. Hence, the existence of Yangian symmetry in five dimensions has the potential to greatly improve our understanding of higher dimensional gauge theories and reveal new dualities in quantum gravity.

\begin{acknowledgments}
We thank Niklas Beisert, Troels Harmark, Neil Lambert, Carlo Meneghelli, Julian Miczajka, Rishi Mouland, and Paul Ryan for useful discussions. AL is supported by the Royal Society via a University Research Fellowship. TO is supported by the STFC studentship ST/S505468/1.
\end{acknowledgments} 

\newpage

\appendix

\section{Indices} \label{indices}
\begin{center}
\begin{tabular}{| c | c | c | c |}
\hline 
Index & Description & Range & $\,$ Metric $\,$ \\
\hline
 & & & \\
$\mu, \nu$ & Spacetime $SO(1,5)$ & $\mu \in \{+,-, i \}$ & $ \eta_{\mu \nu} $ \\
 & & &\\
$i, j$ & Spacial $SO(4)$ & $i \in \{1, \dots, 4\}$ & $\delta_{ij}$\\
 & & &\\
$\alpha, \beta$ & Spinor $Spin(1,5)$ & $\alpha \in \{ 1, \dots, 8 \}$ & $C^{\alpha \beta}$ \\
 & & &\\
$\hat{\alpha}, \hat{\beta}$ & Chiral $Spin(1,5)$ & $\hat{\alpha} \in \{ 1, \dots, 4 \}$ & $\hat{C}^{\hat{\alpha} \hat{\beta}}$ \\
 & & &\\
$\hat{a},\hat{b}$ & Little group $SU(2)$ & $\hat{a} \in \{ 1,2 \}$ & $\epsilon_{\hat{a}\hat{b}}$ \\
 & & &\\
$I, J$ & R-symmetry $SO(5)$ & $I \in \{1, \dots, 5 \}$ & $\delta_{IJ}$ \\
 & & &\\
$A,B$ & R-symmetry $USp(4)$ & $A \in \{1, \dots, 4\}$& $\tilde{\Omega}^{AB}$ \\
 & & &\\
$\hat{A},\hat{B}$ & R-symmetry $U(2)$ & $\hat{A} \in \{1, 2\}$ & $\delta_{\hat{I} \hat{J}}$\\
 & & &\\
$a,b$ & Adjoint index & $a \in \{1, \dots, 50\}$ & $g_{ab}$ \\
 & & &\\
\hline
\end{tabular}
\end{center}
We work in lightcone coordinates, where the Minkowski metric $\eta_{\mu\nu}$ is given by $\eta_{+-} = \eta_{-+} = -1$ and $\eta_{ij}= \delta_{ij}$. We raise and lower little group indices using
\begin{equation}
\epsilon_{\hat{a}\hat{b}}=\left(\begin{array}{cc}
0 & 1\\
-1 & 0
\end{array}\right),\,\,\,\epsilon^{\hat{a}\hat{b}}=\left(\begin{array}{cc}
0 & -1\\
1 & 0
\end{array}\right).
\end{equation} 
Finally, we use the following definition of the 4d self-dual t'Hooft matrices:
\begin{equation}
\eta_{\kappa ij}=\epsilon_{\kappa ij4}+\delta_{\kappa i}\delta_{j4}-\delta_{\kappa j}\delta_{i4},
\end{equation}
where $\kappa\in\left\{ 1,2,3\right\}$ is trivially raised and lowered.
  
\section{Spinor conventions} \label{spinors}
We work with 6d Dirac matrices $\Gamma_\mu$ and 5d Dirac matrices $\tilde{\Gamma}_I$ (associated with the $SO(5)$ R-symmetry) which satisfy 
\begin{align}
&\big \{ \Gamma_{\mu}, \Gamma_{\nu} \big \}  = 2 \eta_{\mu \nu} \mathbb{I}_8, \quad {\Gamma_{\mu}}^{\dagger} = \Gamma_{0} \Gamma_{\mu} \Gamma_{0},  \\
&\big \{ \tilde{\Gamma}_{I}, \tilde{\Gamma}_{J} \big \}  = 2 \delta_{I J} \mathbb{I}_4, \quad {\tilde{\Gamma}_{I}}^{\dagger} = \tilde{\Gamma}_{I}.
\end{align}
The 6d charge conjugation matrix satisfies
\begin{align}
&C^{\dagger}C = \mathbb{I}_8, \quad C^T = C \ \implies \ C^{\dagger} = C^* = C^{-1}, \nonumber \\
&(C\Gamma_{(r)})^T = - t^{\text{6d}}_{r} C \Gamma_{(r)},\nonumber \\
& \text{with} \ t^{\text{6d}}_{0} = t^{\text{6d}}_{3} = -1, \ t^{\text{6d}}_{1} = t^{\text{6d}}_{2} = 1 \ (r \, \text{mod} \, 4) ,
\end{align}
while the $4 \times 4$ orthosymplectic matrix $\tilde{\Omega}_{AB}$ can be considered as a 5d charge conjugation matrix:
\begin{align}
&\tilde{\Omega}^{\dagger} \tilde{\Omega} = \mathbb{I}_4, \quad \tilde{\Omega}^T = - \tilde{\Omega} \ \implies \ \tilde{\Omega}^{\dagger} = - \tilde{\Omega}^* = \tilde{\Omega}^{-1} \nonumber \\
&(\tilde{\Omega} \tilde{\Gamma}_{(r)})^T = - t^{\text{5d}}_r \tilde{\Omega} \tilde{\Gamma}_{(r)}, \nonumber \\
& \text{with} \ t^{\text{5d}}_0 = t^{\text{5d}}_1 = 1, \ t^{\text{5d}}_2 = t^{\text{5d}}_3 = -1 \ (r \, \text{mod} \, 4).
\end{align}
We define the $B$ matrix as $B = i t^{6d}_0 C \Gamma_{0}$, which in our conventions is
\begin{align}
B = -i C \Gamma_{0}.
\end{align}
The 6d spinors are symplectic-Majorana-Weyl and in Minkowski signature obey 
\begin{align}
\label{symplectic}
\chi_A = \tilde{\Omega}_{AB} B^{-1} (\chi_{B})^* = i \tilde{\Omega}_{AB} \Gamma_{0}(C \chi_B)^*.
\end{align}

We also define the matrices $\Sigma, \widetilde{\Sigma}$ in terms of the $\Gamma$ matrices as follows:
\begin{align}
\Sigma_{\mu \hat{\alpha} \hat{\beta}} =  \Gamma_{\mu \hat{\alpha} }^{ \ \ \ \beta} C_{\beta \hat{\beta}} , \quad \widetilde{\Sigma}_{\mu}^{\hat{\alpha} \hat{\beta}} = C^{\hat{\alpha} \beta}  \Gamma_{\mu \beta }^{ \ \ \ \hat{\beta}},
\end{align}
where the range of $\hat{\alpha}$ is $ \{1, \dots, 4 \}$, the first half of $\alpha$. These can be thought of as Clebsch-Gordan coefficients relating two $\bf{4}$'s of $SU(4)$ to a $\bf{6}$. We also define the reduced charge conjugation matrix and its inverse
\begin{align}
\hat{C}^{\hat{\alpha}\hat{\beta}} = C^{\hat{\alpha} (\hat{\beta} +4)}, \quad \hat{C}_{\hat{\alpha} \hat{\beta}} = C_{(\hat{\alpha}+4) \hat{\beta}}.
\end{align}

\section{6d superalgebra} \label{6dalgebra}

The 6d $(2,0)$ superalgebra is $OSp(8|4)$, the bosonic part of which is $SO(6,2) \times SO(5)$. In Minkowski signature we choose conventions where all bosonic generators are anti-hermitian. Their commutation relations are
\begin{align}
  [\hat{M}_{\mu \nu},\hat{M}_{\rho \sigma}] &= \eta_{\nu \rho} \hat{M}_{\mu \sigma} + \eta_{\mu \sigma} \hat{M}_{\nu \rho}\nonumber \\ & \quad - \eta_{\mu \rho} \hat{M}_{\nu\sigma} - \eta_{\nu\sigma} \hat{M}_{\mu\rho} \nonumber  \\
  [\hat{M}_{\mu\nu},\hat{P}_\rho] &= \eta_{\nu \rho} \hat{P}_\mu - \eta_{\mu \rho} \hat{P}_\nu \nonumber  \\
  [\hat{M}_{\mu\nu},\hat{K}_\rho] &= \eta_{\nu \rho} \hat{K}_\mu - \eta_{\mu \rho} \hat{K}_\nu \nonumber \\
  [\hat{D},\hat{P}_\mu] &= - \hat{P}_\mu \nonumber \\
  [\hat{D},\hat{K}_\mu] &= \hat{K}_\mu \nonumber \\
  [\hat{K}_\mu,\hat{P}_\nu] &= 2 \big (  \eta_{\mu\nu} \hat{D} + \hat{M}_{\mu\nu} \big )\nonumber \\
[\hat{R}_{IJ}, \hat{R}_{KL}] &= \delta_{J K} \hat{R}_{I L} + \delta_{I L} \hat{R}_{J K}\nonumber \\  
& \quad - \delta_{I K} \hat{R}_{JL} - \delta_{JL} \hat{R}_{IK}.
\end{align}

The fermionic generators are 6d symplectic-Majorana-Weyl:
\begin{align}
\hat{Q}_{\alpha A} = i \tilde{\Omega}_{AB} (\Gamma_{0})_{\alpha}^{\ \beta} (C\hat{Q}_{B})^{\dagger}_{\beta},
\end{align}
and similar for $\hat{S}$ (again this is in Minkowski signature). Note that the dagger acts by complex conjugation without transposing spinor indices. The $\hat{Q}$ and $\hat{S}$ generators have opposite chirality under $\Gamma_* = \Gamma_{012345}$:
\begin{align}
\label{weyl}
\Gamma_* \hat{Q} = -\hat{Q}, \quad \Gamma_* \hat{S} = \hat{S}.
\end{align}
The commutation relations of the fermionic and bosonic generators are as follows:
\begin{align}
\begin{split}
[\hat{M}_{\mu \nu}, \hat{Q}] &= - \frac{1}{2} \Gamma_{\mu \nu} \hat{Q}, \\
[\hat{K}_{\mu}, \hat{Q}] &= \Gamma_{\mu} \hat{S}, \\
[\hat{D}, \hat{Q}] &= -\frac{1}{2} \hat{Q}, \\
[\hat{R}_{IJ}, \hat{Q}_{A}] &= -\frac{1}{2} (\tilde{\Gamma}_{IJ})_{A}^{\ \ B} \hat{Q}_B,
\end{split}
\begin{split}
[\hat{M}_{\mu \nu}, \hat{S}] &= - \frac{1}{2} \Gamma_{\mu \nu} \hat{S}, \\
[\hat{P}_{\mu},  \hat{S}] &= \Gamma_{\mu} \hat{Q}, \\
[\hat{D}, \hat{S}] &= \frac{1}{2} \hat{S}, \\
[\hat{R}_{IJ}, \hat{S}_{A}] &= -\frac{1}{2} (\tilde{\Gamma}_{IJ})_{A}^{\ \ B} \hat{S}_B,
\end{split}
\end{align}
while the anticommutation relations of the fermionic generators are
\begin{align}
\begin{split}
\{ \hat{Q}_{\alpha A}, \hat{Q}_{\beta B} \} &= 2 \tilde{\Omega}_{AB}(\Pi_- \Gamma_{\mu} C)_{\alpha \beta} \hat{P}^{\mu}, \\ 
\{\hat{S}_{\alpha A},  \hat{S}_{\beta B} \} &= 2\tilde{\Omega}_{AB} (\Pi_+ \Gamma_{\mu}C)_{\alpha \beta} \hat{K}^{\mu}, \\
\{\hat{Q}_{\alpha A}, \hat{S}_{\beta B} \} &= -2(\Pi_- C^{-1})_{\alpha \beta} \big ( (\tilde{\Gamma}_{IJ} \tilde{\Omega}^{-1})_{AB} \hat{R}_{IJ} + \tilde{\Omega}_{AB} \hat{D} \big ) \\ 
&  \quad \, + (\Pi_- \Gamma_{\mu \nu} C^{-1} )_{\alpha \beta} \tilde{\Omega}_{AB} \hat{M}^{\mu \nu},
\end{split}
\end{align}
where $\Pi_{\pm} = \frac{1}{2}(\mathbb{I}_8 \pm \Gamma_{*})$.

\section{5d superalgebra} \label{sec:subalgebra}
The non-vanishing commutation relations of the bosonic generators are
\begin{align}
[M_{i+},P_j]	&=	-\delta_{ij} P_+ & \hspace*{-17pt} - \frac{1}{2}\Omega_{ij} T - \frac{2}{R} \delta_{ij} B&  + \Omega_{ik}\eta^{\kappa}_{jk} C^{\kappa} \nonumber \\
[T,P_-]			&=		-2P_- \quad	&	 [K_+,P_-]	&=	-2T  \nonumber \\
[T,P_i]			&=		-P_i \quad & [P_i,P_j]		&= 	-\Omega_{ij} P_-  \nonumber \\
[T,M_{i+}]		&=		M_{i+} \quad & [M_{i+},M_{j+}]	&=		-\frac{1}{2} \Omega_{ij} K_+  \nonumber \\
[T,K_+]			&=		2K_+ \quad & [C^\kappa, C^\rho]		&= 		- \varepsilon^{\kappa \rho \omega}C^\omega  \nonumber \\
[P_-,M_{i+}]	&=		P_i \quad & [B, X_i]		&= 	\frac{1}{2}R\Omega_{ij}X_j \nonumber \\
[K_+,P_i]		&=		-2M_{i+} \quad & [C^{\kappa}, X_i]		&=	-\frac{1}{2} \eta^{\kappa}_{ij}X_j,
\label{eq: extended su(1,3) algebra}
\end{align}
where $\eta_{ij}^\kappa$ are a basis of 4d self dual 't Hooft matrices, $\kappa \in \{1,2,3\}$, and in the last two expressions we use $X_i$ to denote any operator with a single space index. Any commutation relation not listed is zero, except for the algebra of $R$-symmetry generators which are the same as those in 6d. 

Next consider the fermionic generators. They have the following nontrivial commutation relations with the bosonic generators:
\begin{align}
\! \! \! \! \! [S_{+}, P_-]	&=	- \Gamma_+ Q_{-}   &   [S_{+}, P_i] &= R \Omega_{ij} \Gamma_{+j} \Theta_{-}   \nonumber \\
\! \! \! \! \! [\Theta_{-}, P_i]&= \frac{1}{2R} \Gamma_i Q_{-}  &  [\Theta_{-}, B] &=	-\frac{1}{8}R \Omega_{ij} \Gamma_{ij} \Theta_{-}   \nonumber \\
\! \! \! \! \! [Q_{-}, C^ \kappa]	&=	\frac{1}{8} \eta_{ij}^\kappa \Gamma_{ij} Q_{-}	 & 
\! \! \! \! \! [S_{+}, C^\kappa]			&=	\frac{1}{8}\eta_{ij}^\kappa \Gamma_{ij} S_{+} \nonumber \\
\! \! \! \! \! [Q_{-}, T]			&=	Q_{-}	 & 	[S_{+}, T]		&=	-S_{+} \nonumber \\
\! \! \! \! \! [Q_{-}, M_{i+}]		&=	-R \Omega_{ij}\Gamma_{j} \Theta_{-}	 &	
\! \! \! \! \! [\Theta_{-}, M_{i+}]	&=	\frac{1}{4R}\Gamma_{- i} S_{+} \nonumber \\
\! \! \! \! \! [Q_{-}, K_+]		&= 	- \Gamma_- S_{+}	 & [\Theta_{- A}, R_{IJ}] &=	\frac{1}{2}(\tilde{\Gamma}_{IJ})_{A}^{\ \ B} \Theta_{-B}  \nonumber \\
\! \! \! \! \! [Q_{-A}, R_{IJ}]	&=	\frac{1}{2}(\tilde{\Gamma}_{IJ})_{A}^{\ \ B} Q_{-B}  &	[S_{+A}, R_{IJ}]	&=	\frac{1}{2}(\tilde{\Gamma}_{IJ})_{A}^{\ \ B} S_{+ B}.
\end{align}
They also have the following anticommutation relations:
\begin{align}
\begin{split}
\{ Q_{- \alpha A}, Q_{- \beta B} \} &= - 4 ( \Gamma_- \Pi_- C^{-1})_{\alpha \beta} \tilde{\Omega}_{AB} P_- \\ \{ S_{+ \alpha A},
S_{+ \beta B} \} &= - 4 ( \Gamma_+ \Pi_+ C^{-1})_{\alpha \beta} \tilde{\Omega}_{AB} K_+ \\
\{ Q_{- \alpha A}, \Theta_{- \beta B} \} &= -2 R \Omega_{ij} ( \Gamma_{-} \Gamma_{j}  \Pi_{+} C^{-1})_{\alpha \beta} \tilde{\Omega}_{AB} P_i \\
\{ S_{+ \alpha A}, \Theta_{- \beta B} \} &= -2 R \Omega_{ij} ( \Gamma_{+} \Gamma_{-} \Gamma_{j} \Pi_{+} C^{-1})_{\alpha \beta} \tilde{\Omega}_{AB} M_{i+},\\
\{ \Theta_{- \alpha A}, \Theta_{- \beta B} \} &= -2 (\Gamma_{-} \Pi_{+} C^{-1})_{\alpha \beta} \tilde{\Omega}_{AB} P_+\\  
&  \quad \, + \frac{1}{4} \Omega_{i j}  (\Gamma_{-} \Gamma_{ij} \Pi_{+} C^{-1})_{\alpha \beta} (\tilde{\Gamma}_{IJ}\tilde{\Omega}^{-1})_{AB} R_{IJ}\\
&  \quad \,  - \frac{4}{R}(\Gamma_{-} \Pi_{+} C^{-1})_{\alpha \beta} \tilde{\Omega}_{AB} B,\\
\{ Q_{- \alpha A}, S_{+ \beta B} \} &= 2(\Gamma_{-} \Gamma_{+} \Pi_{+} C^{-1})_{\alpha \beta}\\
&  \quad \, \times \big ( \tilde{\Omega}_{AB} T  - (\tilde{\Gamma}_{IJ}\tilde{\Omega}^{-1})_{AB} R^{IJ} \big )\\
&  \quad \, - ( \Gamma_{ij} \Gamma_{-} \Gamma_{+} \Pi_{+} C^{-1})_{\alpha \beta} \tilde{\Omega}_{AB} \eta^{\kappa}_{ij} C^{\kappa}.
\end{split}
\end{align}
The 5d superalgebra has 50 generators ($26$ bosonic and $24$ fermionic). The bosonic subalgebra generates $U(1) \times SU(1,3) \times SO(5) \, \cong \, U(1) \times SO(6) \times USp(4)$. 


\section{Serre Relations} \label{pseudocode}
In this Appendix, we will describe the method we used to check that the Serre relations are obeyed for the representation constructed in section IV of the main text. For the representation we consider it is sufficient to check the single-site Serre relations, although we also verify them in the multi-site case. In the single-site case the level-1 generators vanish and the Serre relations reduce to
\begin{align}
\sum_{e,g,h, l, m, n} (-1)^{|e| |m| + |h| |n|} \{ J_{(0)}^l, J_{(0)}^m, J_{(0)}^n] f^{a e}_{\ \ l} f^{b g}_{\ \ m} f^{c h}_{\ \ n} f_{e g h} = 0.
\label{singlesiteserre}
\end{align}
We checked this formula using our Mathematica code {\em 5dYangian.nb}, which we include with the submission and outline below.

The generators are encoded as symbolic constants, which we store in Mathematica as an array \verb|J|. The supercommutator is defined by the function \verb|Com[*,*]|, and the commutation relations are entered manually as a lookup table. We illustrate some examples using pseudocode below:
\begin{verbatim}
Com[Kplus, Pminus] :=   -2 T;
Com[Pminus, Kplus] :=     2 T;
Com[P[i_], P[j_]]  := Com[P[i], P[j]]  
                    = -\[CapitalOmega][[i,j]] Pminus;
\end{verbatim}
From this we can calculate the structure constants $f^{ab}_{\ \ c}$ which we we define as \verb|f|:
\begin{verbatim}
f = ParallelTable[
     Coefficient[Com[J[[i]], J[[j]]], J[[l]]], 
  {i, Length[J]},{j, Length[J]}, {l, Length[J]}];
\end{verbatim}

To implement the matrix representation of the algebra defined in section IV of the main text we again use lookup tables and define the function \verb|Mat[*]|, which takes as input a symbol of a generator from \verb|J| and returns a $24\times24$ matrix representative. 
We also define \verb|matJ| which is an array of the matrix representations of the generators of \verb|J|.
From these we can calculate the metric of the representation
\begin{verbatim}
g = ParallelTable[
   Str[matJ[[a]].matJ[[b]]] // Simplify, 
   {a, Length[J]}, {b, Length[J]}];
\end{verbatim}
where \verb|Str[*]| is the supertrace. This allows us to raise and lower the indices of \verb|f| as required in the Serre relations. 

The key object that needs to be calculated is the 6-tensor
\begin{align}
\sum_{e,g,h} f^{a e}_{\ \ l} f^{b g}_{\ \ m} f^{c h}_{\ \ n} f_{e g h},
\end{align}
which is then contracted with a symmetrised triplet of generators according to \eqref{singlesiteserre}. To carry out this sum more efficiently in Mathematica we break it up into contiguous blocks of indicies over which the sign in summation does not change. We calculate these using \verb|Dot| and then add the terms together at the end with appropriate sign. In more detail, decomposing
\begin{align}
\sum_{e,g,h, l, m, n} (-1)^{|e| |m| + |h| |n|} (\dots)
\end{align}
and using the symmetries of $f$ to rearrage the expression in $(\dots)$ such that its indices flow, we find 9 changes of sign:
\begin{align}
\label{eq:lines}
& -\sum_{\substack{s, l, m , n \in \text{all} \\ t, r \in \text{bose} \\  }}
&& + 
\sum_{\substack{s, l, m \in \text{all} \\ r, n \in \text{bose} \\  t \in \text{ferm}}}
&&- 
\sum_{\substack{s, l, m \in \text{all} \\ r,  \in \text{bose} \\  t, n \in \text{ferm}}} \nonumber \\
& + 
\sum_{\substack{s, l, n \in \text{all} \\ t, m \in \text{bose} \\ r \in \text{ferm} }}
&& - 
\sum_{\substack{s, l, \in \text{all} \\ m, n \in \text{bose} \\ r, t \in \text{ferm}}} 
&& + 
\sum_{\substack{s, l, \in \text{all} \\ m \in \text{bose} \\ r, t, n \in \text{ferm}}} \nonumber \\
& - 
\sum_{\substack{s, l, n \in \text{all} \\ t \in \text{bose} \\ r, m \in \text{ferm} }}
&&+
\sum_{\substack{s, l, \in \text{all} \\ n \in \text{bose} \\ m, r, t \in \text{ferm}}} 
&& - 
\sum_{\substack{s, l, \in \text{all} \\ r, t, m, n \in \text{ferm}}}. \nonumber \\
\end{align}
We label these terms based on their position in \eqref{eq:lines}. For instance the first sum on the first line is referred to as \verb|group1line1| . We also flatten the first 3 indices of each of these objects to a single index in order to further improve the efficiency of index contractions. This recasts 6-tensors as 4-tensors with an extended range of the first index. e.g. \verb|group1line1| has dimensions $125000\times 50\times 50\times 50$, as the first three indices originally all had range 50.

Next we create an array of the graded symmetrized triplets of generators:
\begin{verbatim}
GradedSym[A_ . B_ . C_] := (A.B.C + 

     ((-1)^(mag[B] mag[C])) A.C.B +...
\end{verbatim}
where \verb|mag[*]| returns $0$ for bosons and $1$ for fermions. We then define the triplets by
\begin{verbatim}
Trips = Table[GradedSym[
         matJ[[a]].matJ[[b]].matJ[[c]] ]
         , ... ]
\end{verbatim}
and similarly define limits of the indices of these objects to match thoses appearing in \eqref{eq:lines}, e.g \verb|Trips11| is summed with \verb|group1line1|, and hence has a single index of range $125000$. With these building blocks, we can finally define a function that checks the single-site Serre relations in \eqref{singlesiteserre} for a given triplet of generators:
\begin{verbatim}
RHSfunc[a1_, b1_, c1_] :=
        -Trips11.group1line1[[All, a1, b1, c1]]
        +Trips12.group1line2[[All, a1, b1, c1]]
        +...
\end{verbatim}
Using this code, we have checked the Serre relations for over 1000 random examples including cases of each type; boson-boson-boson, boson-boson-fermion etc. 

We have also checked the Serre relations for the case of three sites, which is the minimum number required to see all the generic structure for any number of sites. We can build a three-site realisation by repeated action of the co-product on the level-0 and level-1 generators. In more detail, the Yangian is a Hopf algebra so possesses a map
\begin{align}
\Delta: \mathfrak{g} \longrightarrow \mathfrak{g} \otimes \mathfrak{g}
\end{align}
known as the co-product. Its action on the level-0 and level-1 generators is as follows:
\begin{align}
\label{eq:level1}
\Delta(J_{(0)}^{a}) & = J_{(0)}^a \otimes \mathbb{I} + \mathbb{I} \otimes J_{(0)}^a \nonumber \\
\Delta(J_{(1)}^a)  & =  J_{(1)}^a \otimes \mathbb{I} + \mathbb{I} \otimes J_{(1)}^a +  f_{ b c}^{\ \ a}  J_{(0)}^c \otimes J_{(0)}^b
\end{align}
We can think of each position in this tensor product as a local site which the generators act on, so the above is a two-site system. Since we take the local part of $J_{(1)}$ to vanish, we have 
\begin{align}
\Delta(J_{(1)}^a)  & =  f_{ b c}^{\ \ a}  J_{(0)}^c \otimes J_{(0)}^b.
\end{align}
Repeated action of the coproduct allows us to generate a representation for any number of sites. In particular for three sites we have 
\begin{align}
\label{eq:3sites}
\Delta(\Delta(J_{(0)}^{a})) & = J_{(0)}^a \otimes \mathbb{I} \otimes \mathbb{I} + \mathbb{I} \otimes J_{(0)}^a \otimes \mathbb{I} + \mathbb{I} \otimes \mathbb{I} \otimes J_{(0)}^a \nonumber \\
\Delta(\Delta(J_{(1)}^a))  & = f_{ b c}^{\ \ a} \big( J_{(0)}^c \otimes J_{(0)}^b \otimes \mathbb{I} +  J_{(0)}^c \otimes \mathbb{I} \otimes J_{(0)}^b \nonumber  \\ & \qquad \qquad \qquad \qquad \qquad \quad + \mathbb{I} \otimes J_{(0)}^c \otimes J_{(0)}^b \big).
\end{align}

We can implement this in Mathematica using the tensor product symbol \verb|\[TensorProduct]|. The only modification we need to make is to the \verb|Com| function to tell it how to act on a tensor triple product:
\begin{align}
&[A\otimes B \otimes C, D \otimes E \otimes F \} = \nonumber \\
&\frac{1}{4} (-1)^{(|B| + |C| ) |D| + |C| |E|} \big( [A, D \} \otimes [B, E \} \otimes [C,F\} \nonumber \\
& \qquad \qquad \qquad \qquad \quad \: \; + [A, D \} \otimes \{B, E ] \otimes \{C,F] \nonumber \\ 
& \qquad \qquad \qquad \qquad \quad \: \; + \{A, D ] \otimes [B, E\} \otimes \{C,F]\nonumber \\ 
& \qquad \qquad \qquad \qquad \quad \: \; + \{A, D ] \otimes \{B, E] \otimes [C,F\} \big)
\end{align}
where $[\cdot, \cdot \}$ denotes the supercommutator and $ \{\cdot, \cdot]$, which we call \verb|AntiCom[a_, b_]|, is the superanticommutator. 
We also require the notion of an identity for sites that are acted upon trivially. In the code we denote this as \verb|\[CapitalIota]| and add the rules
\begin{verbatim}
Com[A_, \[CapitalIota]] := 0;
Com[\[CapitalIota], A_] := 0;
AntiCom[A_, \[CapitalIota]] := 2 A;
AntiCom[\[CapitalIota], A_] := 2 A;
AntiCom[\[CapitalIota], \[CapitalIota]] := 
                            2 \[CapitalIota];
\end{verbatim}
We also modify the inbuilt dot function as follows:
\begin{verbatim}
Unprotect[Dot];
Dot[A_, \[CapitalIota]] := A;
Dot[\[CapitalIota], A_] := A;
Dot[\[CapitalIota], \[CapitalIota]] 
                  := \[CapitalIota];
Protect[Dot];
\end{verbatim}

For three sites the left hand side of the Serre relations in (21) of the main text is no longer trivial so we define a function that computes this as well:
\begin{verbatim}
LHSfunc[a1_, b1_, c1_] := 
    Com[J1[[a1]], Expand[Com[J1[[b1]], J0[[c1]]]]]
    + ...
\end{verbatim}
Using all of the building blocks described above we have checked the three-site Serre relations for over 1,000 randomely chosen examples.

\section{Level-1 generators} \label{level1}
The level-1 Yangian generators are bi-local operators constructed from level-0 generators, as defined in (23) of the main text. We present bosonic level-1 generators below:
\begin{widetext}
\begin{align}
\begin{aligned}
P_{-(1)} &= \sum_{u < v}^{N_s} \frac{1}{4} ( P_{- u} T_v - T_u P_{-v})- \frac{1}{4} R^2 \Omega_{ij} P_{i \, u} P_{j \, v} + \frac{1}{8} (\widetilde{\Sigma}_{-})_{\hat{\alpha} \hat{\beta}} \big( Q^{\hat{\alpha}}_{- \hat{A} \, u } Q^{\hat{\beta} \hat{A}}_{- \phantom{\hat{A}} \! \! v} - Q^{\hat{\alpha} \hat{A}}_{- \phantom{\hat{A}} \! \! u} Q^{\hat{\beta}}_{- \hat{A} \, v} \big) \\
P_{i(1)} &= \sum_{u < v}^{N_s} \frac{1}{4} (P_{- \, u} M_{+i \, v} - M_{+i \, u} P_{- \, v}) + \frac{1}{8} ( P_{i \, u} T_{v} - T_u P_{i \,v} ) + \frac{1}{4} \eta^{\kappa}_{ij} (P_{j \, u} C^{\kappa}_{v} - C^{\kappa}_{u} P_{j \, v}) \\
&  \quad -\frac{1}{2}R \Omega_{ij} (P_{j \, u} B_{v} - B_{u} P_{j \, v}) + \frac{1}{4\sqrt{2}} (\widetilde{\Sigma}_i)_{\hat{\alpha} \hat{\beta}} \big( Q^{\hat{\alpha} \hat{A}}_{- \phantom{\hat{A}} \! \! u} \Theta^{\hat{\beta}}_{- \hat{A} \, v}  - Q^{\hat{\alpha}}_{- \hat{A} \, u} \Theta^{\hat{\beta} \hat{A}}_{- \phantom{\hat{A}} \! \! v} - \Theta^{\hat{\alpha}}_{- \hat{A} \, u} Q^{\hat{\beta} \hat{A}}_{- \phantom{\hat{A}} \! \! v} +  \Theta^{\hat{\alpha} \hat{A}}_{- \phantom{\hat{A}} \! \! u} Q^{\hat{\beta}}_{- \hat{A} \, v} \big) \\
M_{+i (1)} &= \sum_{u < v}^{N_s} - \frac{1}{8}(P_{i \, u} K_{+ \, v} - K_{+ \, u} P_{i \, v} ) + \frac{1}{8}( M_{+i \, u} T_v - T_u M_{+i \, v}) + \frac{1}{4} \eta^{\kappa}_{ij} (M_{+ j \, u} C^{\kappa}_v - C^{\kappa}_u M_{+j \, v}) \\
& \quad -\frac{1}{2} R \Omega_{ij} (M_{+j \, u} B_v - B_u M_{+j \, v}) + \frac{1}{8} (\Sigma_i)^{\hat{\alpha}}_{\ \hat{\beta}} \big( S^{\hat{A}}_{+ \hat{\alpha}  \phantom{\hat{i}} \!  u} \Theta^{\hat{\beta}}_{- \hat{A} \, v} - S_{+ \hat{\alpha} \hat{A} \, u} \Theta^{ \hat{\beta} \hat{A} }_{- \phantom{\hat{i}} \! v} +  \Theta^{\hat{\beta}}_{- \hat{A} \, u} S^{\hat{A}}_{+ \hat{\alpha}  \phantom{\hat{i}} \!  v} - \Theta^{ \hat{\beta} \hat{A} }_{- \phantom{\hat{i}} \! u}  S_{+ \hat{\alpha} \hat{A} \, v} \big) \\
C^{\kappa}_{(1)} &= \sum_{u < v}^{N_s} \frac{1}{8}R^2 \Omega_{ik} \eta^{\kappa}_{kj}(M_{+i \, u} P_{j \, v} - P_{i \, u} M_{+j \, v} ) + \frac{1}{2} \varepsilon^{\kappa \rho \omega} C^{\rho}_{u} C^{\omega}_{v} \\
& \quad + \frac{1}{64} \eta^{\kappa}_{ij} (\Sigma_{ij})_{\hat{\alpha}}^{\ \hat{\beta}} \big(Q^{\hat{\alpha} \hat{A}}_{- \phantom{\hat{A}} \! \! u} S_{+ \hat{\beta} \hat{A} \, v} - Q_{- \hat{A} \, u}^{\hat{\alpha}} S_{+ \hat{\beta} \phantom{\hat{A}} \! \! v}^{\hat{A}} + S_{+ \hat{\beta} \hat{A} \, u} Q_{- \phantom{\hat{A}} \! v}^{\hat{\alpha} \hat{A}} - S_{+ \hat{\beta} \phantom{\hat{A}} \! \! u}^{ \hat{A}} Q_{- \hat{A} \, v}^{\hat{\alpha}} \big) \\
B_{(1)} &= \sum_{u < v}^{N_s} - \frac{1}{8} R (P_{i \, u} M_{+i \, v} - M_{+i \, u} P_{i \, v} ) + \frac{1}{8}R (\widetilde{\Sigma}_-)_{\hat{\alpha} \hat{\beta}} \big( \Theta^{\hat{\alpha}}_{- \hat{A} \, u} \Theta^{\hat{\beta} \hat{A}}_{- \phantom{\hat{A}} \! \! v} - \Theta^{\hat{\alpha} \hat{A}}_{- \phantom{\hat{A}} \! \! u} \Theta^{\hat{\beta}}_{- \hat{A} \, v} \big)      \\
T_{(1)} &= \sum_{u < v}^{N_s} -\frac{1}{4}  (P_{- \, u} K_{+ \, v} - K_{+ \, u} P_{- \, v} ) + \frac{1}{4} R^2 \Omega_{i j} (P_{i \, u} M_{+ j \, v} - M_{+i \, u} P_{j \, v} )  \\
& \quad - \frac{1}{8} \big(Q^{\hat{\alpha} \hat{A}}_{- \phantom{\hat{A}} \! \! u} S_{+ \hat{\alpha} \hat{A} \, v} - Q^{\hat{\alpha}}_{- \hat{A} \, u} S^{\hat{A}}_{+  \hat{\alpha} \phantom{\hat{A}}  \! \! v} -  S_{+ \hat{\alpha} \hat{A} \, u} Q^{\hat{\alpha} \hat{A}}_{- \phantom{\hat{A}} \! \! v} + S^{\hat{A}}_{+  \hat{\alpha} \phantom{\hat{A}}  \! \! u} Q^{\hat{\alpha}}_{- \hat{A} \, v} \big)   \\
K_{+(1)} &= \sum_{u < v}^{N_s} \frac{1}{2} R^2 \Omega_{ij} M_{+ i \, u} M_{+j \, v} - \frac{1}{4} ( T_u K_{+ \, v} -K_{+ \, u} T_v) - \frac{1}{8}(\Sigma_+)^{\hat{\alpha} \hat{\beta}} \big( S^{\hat{A}}_{+ \hat{\alpha}  \phantom{\hat{i}} \!  u} S_{+ \hat{\beta} \hat{A} \, v} - S_{+ \hat{\alpha} \hat{A} \, u} S^{\hat{A}}_{+ \hat{\beta}  \phantom{\hat{i}} \!  v} \big), \\
\end{aligned}
\end{align} 
\end{widetext}
where the generators appearing on the right-hand-side of each equation are understood to be level-0 generators.

\end{document}